**Natural intelligence and anthropic reasoning**

Predrag Slijepcevic

Department of Life Sciences

College of Health and Life Sciences

Brunel University London

Uxbridge, UB8 3PH

United Kingdom

e: predrag.slijepcevic@brunel.ac.uk

**Abstract**

This paper aims to justify the concept of natural intelligence – the type of intelligence wider than human intelligence and its derivative, AI. I will argue that the process of life is (i) a cognitive process and (ii) that organisms, from bacteria to animals, are cognitive agents. To justify these arguments, the neural-type intelligence represented by the form of reasoning known as anthropic reasoning will be compared and contrasted with types of intelligence explicated by four disciplines of biology – relational biology, evolutionary epistemology, biosemiotics and the systems view of life – not biased towards neural intelligence. The comparison will be achieved by asking the following questions: 1. Are human observers the only observers within the pool of terrestrial life forms? 2. If not, (a) what's the frequency of non-human observers within the pool of terrestrial life forms; and (b) if all life forms are observers, what's the boundary between the observing and non-observing capacities? 3. Are there true observers within the pool of terrestrial life forms amongst the reference classes of observers that are not human? 4. Do human observers and other observers and true observers share common features? To answer these questions I will rely on a range of established concepts including SETI (search for extraterrestrial intelligence), Fermi's paradox, bacterial cognition, versions of the panspermia theory, as well as some newly introduced concepts including biocivilisations, cognitive universes, and the cognitive multiverse. The key point emerging from the answers is that the process of cognition – the essence of natural intelligence – is a biological universal.

Keywords: natural intelligence, SETI, anthropic reasoning, biosemiotics, evolutionary epistemology, Fermi's paradox.

1. **Introduction**

An average biology student will almost certainly struggle if asked to explain the meaning of the term 'natural intelligence' – the type of intelligence wider than human intelligence and its derivative known as artificial intelligence or AI. This is because biology departments rarely teach the subject of natural intelligence. Most biologists are shy when it comes to using the adjective *intelligent* to describe the behavior of bacteria, archaea, protists, fungi, plants or non-human animals. The concept of natural intelligence is foreign to mainstream biology. From Descartes onwards, all non-human organisms have been considered mere machines, or in the vocabulary of neo-Cartesians such as Richard Dawkins, dumb "lumbering robots" controlled by the selfish genes (Dawkins 1976).

On the other hand, scientists including biologists, are generous when it comes to characterizing human intelligence. The usual argument is that the neural-type intelligence, which peaked in the evolution with *Homo sapiens*, is the superior form of intelligence (Kurzweil 1990), which many believe can only be surpassed by its derivative, AI. Futurists predict the emergence of the AI-based intelligence known variously as the technological singularity (Vinge 1993), superintelligence (Bostrom 2014), machinocene (Price 2016) or post-biological evolution (Dick 2008). On this anthropic scale of intelligence, reserved for humanity and man-made machines, there is little room for other species. Even if there are still unappreciated forms of natural intelligence, it seems likely that the armies of cognitive neuroscientists, evolutionary psychologists, philosophers of mind, AI experts, SETI (Search for Extraterrestrial Intelligence) researchers, engineers, and futurists would miss them because of the anthropic bias. All these experts, many of whom remain Cartesians, fail to appreciate subtle biological phenomena that lead to complex organic forms far more sophisticated than man-made machines can ever be (Elsasser 1987; Rosen 1991).

However, natural intelligence is a welcome and non-controversial subject in certain fields of biology including evolutionary epistemology (Plotkin 1982; Bradie 1986; Plotkin 1987; Gontier 2006), relational biology (Rosen 1985; Louie and Kercel 2007; Louie 2010), the systems view of life (Capra 1996; Capra and Luisi 2014) and biosemiotics (Barbieri 2009; Kull 2015; Kilstrup 2015). In the settings of these disciplines, human-type intelligence becomes only one form of natural intelligence among many other forms. Natural intelligence is equivalent to natural epistemology (Plotkin 1987; Trewavas 2017; Slijepcevic 2018; Calvo et al. 2019) – a biologically neutral position according to which organisms, from bacteria to animals, are cognitive agents and the processes of life and evolution are forms of natural learning (Slijepcevic 2019a).

Yet, versatile forms of natural intelligence, from bacteria to ecosystems (Slijepcevic 2018), lack properly defined unifying principles. As a result, mainstream science ignores the subject of natural intelligence and often views it as a controversial and risky subject, not worthy of experimental testing (Chamovitz 2018). By contrast, the textbooks, scientific journals and popular books that deal with human-type intelligence and AI are parts of the scientific mainstream even though they lack a proper biological grounding.

The purpose of this study is to justify the concept of natural intelligence by comparing arguments from four disciplines of biology not biased towards neural intelligence – relational biology, evolutionary epistemology, biosemiotics and the systems view of life – with arguments that are rooted in and biased towards the neural-type intelligence, such as those used in the Anthropic Principle (AP) reasoning or anthropic reasoning (Carter 1974; Barrow and Tipler 1996). In brief, the universe appears to be fine-tuned for intelligent life and the emergence of observers – if cosmological constants are not exactly as they are, there would be no observers. The AP debate centers around two arguments: (i) the universe is intentionally created for us as intelligent observers and (ii) intelligent life is the result of improbable events that would not be replicated in other universes. The first option is consistent with the existence of an external "creator", either God (an option that is discarded as unscientific), or the possibility that life on Earth is seeded from other cosmic locations as argued by the proponents of the directed panspermia hypothesis (Crick and Orgel, 1973). The second option implies the existence of many universes (multiverse), one (or only a few) of which (including our universe) contains life (Tegmark 2009). However, there is another option that may be called the life principle (Davies 2003). Given the abundance of the molecular material for life and the ubiquity of stable stars and habitable planets, the emergence of life is a 'cosmic imperative' (de Duve 2011). The life principle is consistent with the concepts of biocentrism (Lanza and Berman 2009), biological determinism (Davies 2003; De Duve 2011) and some forms of the panspermia hypothesis, such as the Hoyle-Wickramasinghe thesis (e.g. Hoyle and Wickramasinghe 1981) and transpermia (Davies 2003). My intention in this study is to use anthropic reasoning to arrive at the position, broadly consistent with life principle, that natural intelligence is a biological universal – organisms from bacteria to animals are cognitive agents – and that the human-type intelligence is only a fraction in the wide spectrum of natural intelligence.

    Here is a brief sketch of anthropic reasoning as derived from Bostrom (2002). The process of sampling the world by which we, as observers, interact with the world is limited by the inability of our observational faculties to capture the totality of information about the world. The consequence of these relationships between observers and the world are observation selection effects – we are necessarily handicapped as observers (Maor et al. 2008). This handicap is frequently dubbed anthropic bias. The function of anthropic reasoning, on the other hand, is to attempt to eliminate various forms of anthropic biases.

This can be achieved by pushing the borders of human understanding of the world to its limits, by combining the cutting-edge scientific evidence with the type of argumentation necessary to avoid pitfalls of the sampling bias. Thus, anthropic reasoning seems to be a suitable method for reducing anthropic bias usually associated with the interpretation of natural intelligence.

I start by giving a brief overview of AP and associated concepts from the perspective of biology (Section 2). I then focus on the concept of observation as understood by AP on one side, and relational biology, evolutionary epistemology, biosemiotics and the systems view of life on the other (Sections 3 & 4). This comparison uncovers fundamental differences between the two sides and paves the way for applying anthropic reasoning further through asking questions aimed at understanding the process of observation and true observers (Section 5). I conclude with the analysis of AP and natural intelligence from the new angle which is not biased towards the neural-type intelligence (Section 6).

## 2. A brief overview of AP and associated concepts

The AP concept was developed by astrophysicists and philosophers as an analogy to the Copernican or mediocrity principle, which states that humans are not expected to occupy a privileged position in the universe (Carter 1974). By contrast, AP is consistent with the notion that the position of *Homo sapiens* is to some extent privileged: we are a species capable of (i) cognition, which includes observation at the scale of the universe (observable universe) and (ii) interpretation of our position in the universe. AP is thus an argument against the Copernican principle. Given that biology is neutral about the position of individual species relative to the rest of nature, AP is criticized by some biologists (e.g. Barash, 2018)[1]. However, this does not mean that AP is without any value to biology. Here is a sketch of AP adapted to biology.

---

[1] There are critics of AP among physicists as well. For example, Smolin (2004) thinks that the AP concept is not scientific because it cannot be falsified. Also, Maor et al. (2008) suggested that "[T]he anthropic principle is however based fundamentally on ignorance rather than knowledge."

Our sensorium registers only a subset of signals/stimuli available in the external world, leading to observation selection effects. For example, we tend to view animals and people as independent individuals, even though they are composed of numerous parts that once possessed individuality. As a result, artists typically depict people as absolute individuals. Lisa Gherardini, used by Leonardo da Vinci (1452-1519) as a model for the *Mona Lisa*, exists in the mind of the artist, and the minds of the observers of *Mona Lisa*, as an undisputed individual. By contrast, a counterintuitive artistic vision of Giuseppe Arcimboldo (1527-1593) depicts individuals as composites (Slijepcevic 2019b) (Figure 1). Unlike *Mona Lisa*, a rigid biological singleton, *Flora* is a composite organism formed by the merger of formerly independent organisms (Figure 1).

From the perspective of modern biology, *Flora* is a more authentic portrait than the *Mona Lisa*. Human bodies are complex ecological collectives consisting of 37 trillion human cells (themselves collectives formed by the microbial mergers) and 400 trillion microbes which together form symbiotic partnerships known as holobionts (Margulis 1993; Zilber-Rosenberg and Rosenberg 2008). Apart from the most basic microbes, bacteria and archaea, which are the only true biological singletons, all other organisms - protists, fungi, plants, and animals - are archaeal-bacterial mergers and thus composites. Absolute individuals do not exist above the level of single-cell prokaryotic microbes. On this understanding, life is an organic conglomerate of transient composite forms. By the same token, evolution is the process of the change of the composites.

Observation selection effects in this case (Fig 1), represent the difference between the intuitive artistic vision of Leonardo da Vinci formed based on insufficient evidence about the true structure of human bodies, and the counter-intuitive artistic vision of Arcimboldo, which, though metaphoric, is in some ways more accurate because the unusual artistic imagination happens to be in line with the modern scientific evidence.

Thus, observation selection effects are the consequence of a specific epistemic position of the human observer relative to the world. This interplay between our capacity to observe the world and the evidence we use to justify the accuracy of our observations is called anthropic reasoning (Bostrom 2002). This type of reasoning stems from the AP concept (Carter 1974) which implies that our position in the universe is privileged to some extent "in so much as special conditions are necessary for our very existence" (Carter 1983). Carter (1974) articulated two forms of AP, known as weak AP (WAP) and strong AP (SAP). WAP states that our position in the universe is privileged because our existence is dominated by our capacity to observe the surrounding world and the universe at large. On the other hand, SAP includes an imperative that the universe *must* allow the emergence of observers at some stage in its development. Barrow and Tipler (1986) formulated their versions of WAP and SAP, which deviate from Carter's by allowing a greater degree of teleology. They also formulated a final AP (FAP), which states that an information processing capacity must emerge in the universe and once it emerges it will never die out. Finally, Wheeler (1994) argued for Participatory Anthropic Principle (PAP) according to which observers are necessary to bring the universe into existence.

In spite of differences between various forms of AP, they share a common feature. All forms of AP assume that the minimum requirement for the true observing capacity is the human-type intelligence (neural intelligence), which eventually peaks in science and technology. Thus, AP holds that the observers must possess, at the minimum, the human-type intelligence plus techno-science, or techno-science-like method, as a form of knowledge acquisition. If we accept this assumption, three types of observers are possible in the universe (i) humanity bound to Earth (and on the verge of the cosmic adventure) possessing techno-science at the present state of development as a form of knowledge acquisition, (ii) humanity-like civilisations living somewhere else in the universe, and (iii) civilisations with the observing capacities superior to the capacities of human or human-like civilisations, living somewhere else in the universe[2]. From this, it also follows that AP assumes two distinct territories of life, as we know it on the planet Earth: (i) intelligent life represented by *Homo sapiens* and its technology in the form of techno-science and (ii) all other forms of life (microorganisms, plants, and animals) considered either non-intelligent or not intelligent enough from the perspective of observing capacities. This position is apparent, for example, in Drake's equation (Drake 1961) devised to estimate the number of extraterrestrial civilisations in the Milky Way galaxy

$$N = R * f_p \, n_e \, f_l \, f_i \, f_c \, L$$

where:

$N$ = the number of civilisations in our galaxy with which we can communicate

$R*$ = average rate of star formation in our galaxy

$f_p$ = fraction of stars with surrounding planets

$n_e$ = number of planets that can support life

$f_l$ = fraction of planets from the $n_e$ pool that develop life

$f_i$ = fraction of planets with life that develop intelligent life

$f_c$ = fraction of civilisations that develop technology for emitting signals in space

$L$ = the length of time during which signals are emitted.

---

[2] Options (ii) and (iii) also imply that these civilisations might have existed in the past and might now be extinct. Option (iii) is not incompatible with some form of post-biological evolution.

Three components in the equation - $f_l$, $f_i$, and $f_c$ - deal with the distinction between the life-developing and intelligence- or technology-developing potentialities of civilisations.

However, I will argue that the key AP assumption is wrong. I define the key AP assumption as follows:

**The human-type intelligence (neural intelligence) and humanity-type civilisation supported by techno-science, is the minimal type intelligence/civilisation capable of the true observation at the cosmic scale.**

I will also argue that there are no fundamental differences between the observing capacities of, for example, bacteria and *Homo sapiens*. Intelligence emerges not only in the case of interacting neural cells but also in the case of interacting bacteria that turn their colonies into brain-like entities (Ben-Jacob 2009; Slijepcevic 2018). My argument is rooted in the notion that the process of life is inherently an observation-like process whereby all organisms are cognitive agents and the process of evolution is a cognitive process (Maturana and Varela 1980). This view is consistent with different versions of life principle including biological determinism (Davies 2003), biocentrism (Lanza and Berman 2009), cosmic imperative (De Duve 2011) and different forms of the panspermia hypothesis (Hoyle and Wickramasinghe 1981; Davies 2003). In particular, the notion of organisms as cognitive agents capable of sensing and processing environmental stimuli or biological information (Slijepcevic 2019a) is consistent with the informational view of life implicit in the concept of biological determinism (Davies 2003).

3. **Arguing against the key AP assumption**

The motivation behind developing an argument against the key AP assumption is in line with anthropic reasoning - the purpose of anthropic reasoning is to reduce anthropic bias.

For the sake of argument, let us assume that Fermi's paradox, or the problem of "great silence"[3], maybe a source of anthropic bias. The expectation is, in line with the astrophysical arguments based on the size and age of our galaxy, that humanity-type civilisations, with developed technologies for cosmic communication and travel, should exist and yet there is no contact with them in spite of SETI efforts (Jones 1985). We can argue further, based on the above assumption, along the following lines. Given that several decades have passed since the establishment of the SETI programme, at the heart of which is the AP assumption, and yet not a single piece of evidence was obtained to support SETI, an alternative hypothesis must be considered[4]. This alternative hypothesis is directed neither against SETI, nor concepts such as Fermi's paradox. Instead, the alternative hypothesis, the main aim of which is to advance anthropic reasoning through reducing the anthropic bias by being more inclusive of the new evidence from biology, may also be beneficial to the SETI programme - it may broaden its narrow scope limited to the neural-type intelligence.

To advance the case against the AP assumption I will proceed as follows. The AP assumption can be shortened into the following statement: **There are no true observers in the universe below the human-type observers.** The first task in developing the counter-argument is to define (i) the process of observation and (ii) the notion of true observers. To achieve this we can rely on Brandon Carter's recognition, long after he formulated AP, that the term "cognition principle" (Carter 1983) may have been less problematic than the somewhat inappropriate term AP.

---

[3] The origin of Fermi's paradox is a famous informal lunch at the Los Alamos laboratory in the spring of 1950, attended by Enrico Fermi and his three physicist colleagues. After a quick calculation, Fermi concluded that there was enough time for extraterrestrial civilisations to visit the Solar system thousands of times. He asked a famous question "Where are they?" and concluded, rather pessimistically, that we are alone and faced with the great cosmic silence.

[4] The SETI timeline (60 years or so) is too short for success. However, the alternative hypothesis can help with eliminating the potential timeline problem. For example, the alternative hypothesis may yield a scenario that is not dependent on the rather short SETI timeline (in the order much lower than the human evolutionary timeline).

Therefore, substituting the word *anthropic* with the word *cognition* (AP becomes CP) provides a shortcut towards defining the concepts of observation and true observers. The process of observation may be interpreted as a cognitive process. It consists of three components: the knowing subject (humanity), the object to be known (the world or the universe) and the pool of knowledge used to justify our position in the world (technoscience at the present state of human development; mythologies, religions, and pre-science philosophy in the past) (Fleck 1981). From the perspective of AP, humanity self-selects[5] evidence from the existing pool of knowledge, best suited to describe our position in the world/universe, in line with the prevailing collective opinion. The AP position is consistent with the view that the cognitive capacity, as described above, exist only in one species on Earth, *Homo sapiens*.

This leads us to consideration of the concept of true observers. The SETI programme implies that true observers must be human-like observers: those observers that can (i) perceive local and cosmic-level information through natural sensorium and man-made technologies, or equivalents, (ii) reflect on the sensory inputs with the help of the accumulated explanatory apparatus (a techno-science like method), and (iii) direct its reaction on the cosmic scale through, for example, emitting and detecting signals in space using appropriate technologies with the hope of establishing contacts with similar non-local civilisations that share temporal and spatial coordinates within the range of the signal

---

[5] Both Carter (1983) and Bostrom (2002) advocate SSA – self-selecting assumption

(Ćirković and Vukotić 2013).[6] Again, the AP position is clear. Only one species (or reference class of observers) on Earth has this capacity, *Homo sapiens*.

To start developing the counter-argument one can ask questions aimed at probing the applicability of the observation process and the concept of true observers on reference classes other than *Homo sapiens*. Here is the list of four such questions.

- Are human observers the only observers within the pool of terrestrial life forms? (Q1)

- If not, (a) what's the frequency of non-human observers within the pool of terrestrial life forms (some life forms, or all life forms); and (b) if all life forms are observers, what's the boundary between the observing and non-observing capacities? (Q2)

- Are there true observers within the pool of terrestrial life forms amongst the reference classes of observers that are not human? (Q3)

- Do human observers and other observers and true observers share common features? (Q4)

To formulate answers to Q1-4 I need to outline positions of the four disciplines of biology, relational biology, evolutionary epistemology, biosemiotics and the systems view of life, about how they interpret the process of cognition. A feature these four disciplines of biology share is the independence from the neural-type intelligence when it comes to interpreting cognition (see below). Only when the positions of these four disciplines of biology are stated, I can proceed with answering Q1-4.

## 4. The concept of cognition from the perspective of relational biology, evolutionary epistemology, biosemiotics, and the systems view of life

---

[6] The form of SETI described here also incorporates a segment known as the active SETI or METI (Messaging to Extra-Terrestrial Intelligence) in section iii.

What follows is a summary of theoretical bases for each discipline of biology. For a more detailed exposition of each topic, readers are referred to the essential set of references used in the text.

**4.1 Relational biology**

The most prominent proponent of the school of relational biology, originally established by a physicist Nicholas Rashevsky, was his Ph.D. student, Robert Rosen (Louie and Kercel 2007; Louie 2010). To explain how science works, Rosen developed the concept of a modelling relationship between the natural system (NS) and the formal system (FS) (Fig 2 A). From the perspective of the human understanding of it, the world is divided into the self (individual and collective humanity) and its ambience (the rest of the world) (Rosen 1991). The structure of the world is such that there is a congruence between the self and its ambience. For example, the self (NS) internalizes the ambience through its model of it (FS), based on language and mathematics. The key thing in the modelling relationship is the concept of information which in Rosen's diagram (Fig 2 A) is represented by arrows indicating encoding and decoding. Rosen used the mathematical category theory to map the relationship between NS and FS. In this form of mapping, encoding represents measurements or abstractions. All measurements are generated through our senses or technological extensions of our senses (scientific instruments). Thus, the abstraction becomes a form of internalization of the ambience by the self through the scientific analysis of entities in the observable world. Decoding, on the other hand, represents an "operator" that makes changes either in NS or FS to test the modelling relationship. Rosen's category-theoretic mapping revealed that there is commutativity between the properties of NS such as causality, and the mathematical properties of FS including entailment (implications of the model, Fig 2 A).

At the heart of Rosen's model is an attempt to explain the epistemological relationship between any natural entity (NS) and knowledge of it (FS) (Fig 2 A). Rosen postulated that the entire nature models itself and it is casually entailed in the form of a modelling relationship presented in Fig 2 A. In other words, science is humanity's way of modelling nature. Similarly to humans, all other species use species-specific epistemological methods to produce modelling relationships.

To explain the universality of Rosen's model Kineman (2007; 2011) and Kineman et al. (2007) introduced biological structure and function into the model (Fig 2 B). The structure and function are the emergent properties of the modelling relationship resulting from the original Rosen's diagram (Fig 2 B). The structure is represented by natural measurements or abstractions executed by the sensorium of each species. The "operator" behind the process of decoding in the original diagram (Fig 2 B), when applied to the entire natural world, becomes the process of the epistemological search for the biological function. The entire process is tested by the filter of natural selection - the filter tests various structure-function forms that emerge from the process of biological abstraction leading to functions such as vision, flight, natural computation, etc. Thus, the empirical world is emerging from the epistemological-ontological unity implicit in Rosen's model. The consequence is that all biological systems, from bacteria to ecosystems, are anticipatory systems that contain internal predictive models of themselves and their environments (Rosen 1985).

**4.2 Evolutionary epistemology**

The programme of research in evolutionary epistemology (Bradie 1986; Plotkin 1987; Gontier 2006) was initiated by Campbell (1960; 1974). The key assumption behind the branch of evolutionary epistemology known as EEM (evolutionary epistemology mechanisms) was that animals are true learners or "knowers" (Plotkin 1982). The original position of EEM was recently revised to take account of the cognitive abilities of all forms of life including prokaryotic microorganisms (bacteria and archaea), eukaryotic microorganisms (protists), fungi (single- and multi-cellular), plants and animals (Slijepcevic 2018; 2019a). This more recent version of EEM reflects better the original principles formulated by Plotkin (1982): (i) living systems are knowledge systems, (ii) evolution is the process of gaining knowledge and (iii) there are features shared by all forms of knowledge gain.

In brief, all organisms, from bacteria to animals, acquire knowledge about their environments through the process of natural learning based on the universal algorithm. The driving force behind the algorithm is the concept of biological information. In the context of biology, information is a purely relational concept. Any form of information becomes actualized only when there is a cognitive system capable of utilizing it. Information, when non-utilized by a cognitive system, exists only as potential information ($I_p$) (Corning 2007). For example, the presence of sugar molecules in a watery solution lacking cognitive systems such as bacteria may be interpreted as a passive form of information or $I_p$ (Slijepcevic 2019a). Once the bacteria become part of the watery environment containing dissolved sugar, they detect the sugar with its sensory apparatus. The passive form of information, or $I_p$, once detected by the bacterial sensory apparatus, becomes actualized and turns into control information ($I_c$) (Corning 2007; Slijepcevic 2019a). This triggers the algorithm for natural learning summarized by the acronym IGPT (information-gain-process-translation). The whole process may be expressed as:

$$I_p \rightarrow I_c \rightarrow IG \rightarrow IP \rightarrow IT$$

where

- IG (Information Gain) = information gathering about the environment by the biological system using its sensory-motor apparatus;
- IP = processing that information by the internal structure of the biological system (natural computation);
- IT = translation of the processes behind IG and IP into structural and behavioural changes of the biological system (Slijepcevic 2019a).

The above algorithm is best viewed as a summary of the process behind biological adaptations. It should be viewed as a descriptor for the sequence of events behind the adaptive process, rather than an algorithm in the truly mathematical sense. In brief, the relationship between organisms and their environments starts with challenges posed by the environment to the integrity of the organism. The organism responds by a complex set of adaptations (IGPT) the aim of which is to solve the problems initiated by the organism-environment interactions. Thus, adaptations incorporate into themselves those aspects of the environment reflecting a particular problem. As a result, organisms behave as cognitive agents, rather than passive objects shaped by the interactions between the genes and the environment. This also means that adaptations are the result of the epistemic process that incorporates cognitive methods including sensing/perception, memory, communication, and decision making, and forms of inheritance including genetic, epigenetic, transgenerational, ecological, psychological inheritances and the inheritance of genomes through predation (Slijepcevic 2018; 2019a).

From the perspective of evolutionary epistemology, adaptation is a natural learning process employed by organisms as cognitive agents. The environmental features represented by $I_p \rightarrow I_c$ are internalized by cognitive agents in a multi-stage IGPT process. $I_p$ and $I_c$ represent

(i) a form of "glue" that holds together organism-environment interactions and (ii) a guiding principle behind natural learning.

## 4.3 Biosemiotics

The concept of semiosis, or sign utilization, has a long history. It started with two opposing schools. On the one hand, a linguist Ferdinand de Saussure (1972) used the term semiology to study signs within the subject of psychology. In his interpretation, the sign was a dual entity consisting of a signifier and a signified. On the other hand, Charles Saunders Peirce (1906), a pragmatist philosopher, invented the triadic concept of sign consisting of a sign vehicle, an object, and an interpretant. This triadic concept is accepted by modern biosemiotics. However, the process of turning the concept of semiosis, which was at best a combination of disparate and frequently contrasting philosophical, linguistic and psychological concepts in the 1960s, into a fully-fledged scientific discipline of modern biosemiotics, included several steps (Barbieri 2009).

The discovery of the genetic code pointed to the existence of "[a] language much older than hieroglyphics, a language as old as life itself … its letters are invisible and its words are buried in the cells of our bodies" (Beadle and Beadle 1966). As soon as it was discovered, the genetic code was interpreted as a form of language used by the cell, which in turn becomes a semiotic system. In this semiotic system, signs or symbols are required for various cellular functions. Pattee (1968) articulated the idea that the cell is controlled by the symbols. By incorporating a range of ideas into semiosis including molecular biology, the genetic code, and Von Neuman self-replicating automata, Pattee argued that "life is matter controlled by symbols" (Pattee 2008).

This paved the way for the emergence of zoosemiotics, the proposal that animal communication is based on signs (Sebeok 1972). The reasoning behind zoosemiotics was largely based on the writings of Jakob von Uexküll. This almost forgotten biologist provided a large body of evidence for the existence of semiosis in the world of animals (Uexküll 2010). Additionally, Thomas Seboek, with the help of collaborators including Prodi (1988) and Krampen (1981), and relying on the evidence generated by other scientists (Sonea and Panisset 1983; Sonea 1987; 1988), articulated the view that a primitive form of semiosis or protosemiosis exists in the world of microorganisms and plants. All this enabled the birth of modern biosemiotics based on two key principles (Barbieri 2009). First, "life and semiosis are coextensive". Biosemiotics is what makes animate matter different from the inanimate equivalent. Second, the existence of signs, meaning, and codes separates biosemiotics from "intelligent design" and doctrines which assume that the origin of life has supernatural roots.

The consequence of biosemiotics is the emergence of the concepts of semiotic scaffold and the semiosphere. The semiotic scaffold is defined as the network of semiotic interactions that permeates the entire nature: the web of sensing, interpreting and coordinating social interactions between organisms of the same species and organisms of different species, through various forms of cross-kingdom communication (Hoffmeyer 2015). The semiosphere, on the other hand, is "[a] sphere like the atmosphere, hydro-sphere, or biosphere. It permeates these spheres from their innermost to outermost reaches and consists of communication: sound, scent, movement, colors, forms, electrical fields, various waves, chemical signals, touch, and so forth—in short, the signs of life" (Hoffmeyer 1996, vi).

**4.4 Systems view of life**

The emergence of systems theory (Bertalanffy 1968), cybernetics (Wiener 1948), information theory (Shannon and Weaver 1949) and complexity theory (Mandelbrot 1983; Nicolis and Prigogine 1989) greatly influenced some of the basic concepts in biology. As a result, living systems are viewed as open thermodynamic systems, far from equilibrium, that constantly exchange matter, energy, and information with their surroundings. The key feature of living systems is not their composition, the nature of chemical constituents that make them up, or matter, but rather the pattern in which the matter is organized to produce various organismal forms (Capra and Luisi 2014). Systems theorists and cyberneticists identified a common pattern of organization that typifies all living systems – the network pattern. Some authors call it reticulate evolution (Gontier 2015). The entire biosphere is a giant network consisting of intertwined webs or networks nesting within the larger networks. The key property of any network is non-linearity – the pattern of organization within the network goes in all directions. The relationship between parts of the network become non-linear. This is a consequence of the fact that as a message, or information, travels along the network pattern it may take a cyclical path leading to the establishment of a feedback loop. Indeed, the organization of living systems from cells to societies is replete with feedback loops, which eventually enable the living systems, including the biosphere at large, to self-regulate.

This system's thinking resulted in two important concepts in biology, autopoiesis and embodied cognition, developed by Maturana and Varela (1980). The key feature of the autopoiesis concept is self-organization or self-making (auto – self; poiesis – making). According to Maturana and Varela every organism, from single-cell microbes to complex multicellular animals, is an autopoietic unit – a system that sustains itself due to the network pattern of organization, which allows constant self-regeneration within the boundary that separates the autopoietic unit from its environment. However, autopoietic units are never truly separated from the environment. There is a structural coupling between the autopoietic unit and its environment. For example, interactions between a bacterial colony and its environment containing antibiotics, lead to structural changes in the bacteria, such as the emergence of antibiotic resistance, and also the structural changes in the environment, such as the sensitivity of susceptible organisms to bacterial infections. This eventually leads to the emergence of bio-entropy, the formation of new patterns whereby the waste created by one living system becomes a useful metabolite for another. This new pattern integrates into the giant network pattern of the biosphere that houses all living systems.

The nature of interactions between organisms and their environments is cognitive - the mind-like or brain-like (Maturana and Varela 1980). Any living organism, irrespective of whether that is a bacterium or an elephant, decides autonomously, through its sensorium faced by various constraints, whether to notice a stimulus in the environment and whether to react to it. Noticing and reacting to the stimulus leads to structural changes within the organism and within its environment. Through these structural changes organisms "bring forth a world", based on their own decisions which stimuli to notice and react to. As Capra and Luisi (2014) suggested: "[C]ognition, then, is not a representation of an independently existing world but rather a continual bringing forth a world through the process of living."

Apart from Maturana and Varela, two more thinkers are worth mentioning as important proponents of the systems view of life, Gregory Bateson and Lynn Margulis. Similarly to Maturana and Varela, Bateson was interested in the pattern of organization in living systems and its underlying principles. He thought that these principles are cognitive principles and used the term "mind" to describe them, but without theological implications. According to Bateson "Mind is the essence of being alive" (cited in Capra and Luisi 2014, 253).

Bateson (1979, p92) proposed six criteria that any system must satisfy to be classified as the mind-like system[7]. Margulis is credited with providing critical evidence for the symbiotic mergers of single-cell prokaryotic microbes, bacteria, and archaea, into more complex eukaryotic cells. Furthermore, Margulis (2004) developed the serial endosymbiotic theory which is in line with the reticulate organization of the biosphere. She suggested that the biosphere is not an organism, because organisms cannot recycle their waste. The biosphere is best viewed as a supersystem dominated by the principles of bio-entropy and capable of self-regulating (Lovelock and Margulis 1974).

## 5. Answering Q1-Q4

### 5.1 Q1

If we (i) accept that the concept of observation is similar or equivalent to the concept of cognition (Carter 1983), (ii) acknowledge that the process of cognition involves three components (knowing subject, object to be known and some form of natural epistemology, or knowledge, as an interface between the two) (Figure 3), and (iii) take into account evidence from the four disciplines of biology (section 4; Figure 3), according to which the process of cognition is not limited to the neural-type intelligence, it becomes possible to argue that human observers are not the only observers within the pool of terrestrial life forms (Figure 3). Even though the four disciplines of biology have independent research programmes they show common elements. For example, they interpret organisms as autonomous natural agents structurally coupled to their environments through species-specific sensoria which become tool-kits in the process of cognition (Table 1). The coupling is driven by the capacity of natural agents to read and process environmental stimuli or natural information (Table 1). This position is consistent with the informational view of life implicit in biological determinism (Davies, 2003) and the increasingly cognitivist interpretation of biology (e.g. Shapiro 2007; Ben-Jacob 2009). Therefore, the answer to Q1 is negative (Table 2).

---

[7] Six criteria are: (1) mind is an aggregate of interacting parts, (2) interaction between parts is triggered by a difference (biological information), mental process (3) requires energy and (4) circular chain of determination, (5) the effects of difference (biological information) are transforms of preceding events and (6) the hierarchy of transformations discloses the hierarchy of logical types.

**5.2 Q2**

The answer to Q2a follows logically from Fig 3 and Table 1. Given that all organisms are natural agents with cognitive faculties (Fig 3; Table 1) all forms of life can be considered forms of observers (Table 2). By this logic, the first forms of life, bacteria, and archaea are the most fundamental observers. All other observers are derived from them through mergers of simpler observing units into more complex ones, also known as the serial endosymbiosis theory (Margulis 2004). This is visible from Fig 1 – the human body is an ecological collective formed by the process of symbiogenesis.

A range of studies carried out in the last two decades, independently of the four disciplines of biology, confirm that the process of cognition is not confined to higher animals, but it is present in microorganisms and plants. For example, bacterial natural sensorium has been characterized as a "bacterial cognitive tool-kit" (Lyon 2015; 2017). The capacity of individual bacteria to establish communication with each other through the bacterial chemical language enables the emergence of intentionality on the part of bacterial colonies which become the brain-like entities (Ben-Jacob 2009). Cognitive faculties of plants have also been documented in spite of objections from some biologists (Trewavas 2017). The biosphere may be viewed as a super-system driven and regulated by the process of cognition. Thus, Darwin's thought, "The difference in mind between man and the higher animals, great as it is, certainly is one of degree and not of kind", can be extended to all life forms. This is in line with Bateson's thought that "Mind is the essence of being alive" (cited in Capra and Luisi, 2014, 253).

Q2b is related to defining the boundary between observing and non-observing biogenic forms. Some biogenic forms must be transitional forms – those forms that have proto-observing or proto-life capacities. Biogenic forms that lack the capacity of observation are viruses. Yet they are the most abundant biogenic forms on Earth (Moelling and Broecker 2019). Viruses cannot be characterized as autonomous agents or cognitive agents, in the same sense as bacteria. Microbiologists usually classify them as obligate parasites that require living cells for their propagation. However, the evidence is emerging that viruses have proto-cognitive faculties such as communication (Erez et al. 2017). Thus, viruses are half-alive: proto-observers or proto-organisms, as shown in Figure 4, in the context of their genome organization relative to bacteria and eukaryotes (Table 2).

The answer to Q2b is that the boundary between observing and non-observing capacities is represented by the proto-observing structures such as viruses (Fig 4; Table 2). The answer is in line with the continuity thesis – the emergence of life is an integral process of the evolution of cosmos (de Duve 2011). The key question then becomes: when exactly proto-observers, such as viruses, emerged in the cosmic history (Fig 5)?

The temporal scale of the cosmos is 13.8 billion years (BY). Life, as we know it, have been existing for at least 3.8 BY. The proto-observers must have emerged at any time between the point at which the process of creation of chemical elements required for biogenic structures was completed (point 1) and the point at which life emerged on Earth in the form of first fully functioning organisms or observers (point 2) (Fig 5). The time-scale between points 1 and 2 is short in the case of the most prevalent hypothesis, that life emerges through the process of abiogenesis. For example, the cosmic imperative thesis (de Duve 2011) is based on the assumption that principles of life are inherent in the laws of physics and chemistry – abiogenesis may be widespread in the cosmos. This means that life will emerge spontaneously at an Earth-like planet. Given that the first planets were formed much earlier than Earth (Lineweaver 2001), it follows that life, in the form of first fully functioning observers such as bacteria-like organisms, might have emerged early in the cosmic history at some first-generation Earth-like planet, following the emergence of the Earth-equivalents of the RNA world and viruses as proto-observers. According to this view, the emergence of life on Earth is just a local cosmic event. As currently understood, the planet Earth was formed 4.5 BYA, leading soon after that to the emergence of the Earth-bound 'RNA world' and viruses as proto-observers, 4.1 BYA – 3.8 BYA, paving the way for the emergence of first living organisms or observers.

Another possibility consistent with the universe-wide abiogenesis process is the transpermia hypothesis. For example, life could have emerged through the process of abiogenesis on a planet like Mars, and then transferred to Earth through the rocky Mars ejecta containing microbes (Davies 2003). It is, therefore, possible that proto-observers such as viruses originated on Mars and then transferred to Earth. The timescale for the emergence of proto-observers, in this case, would be similar to the timescale applicable to the cosmic imperative thesis (Fig 5).

However, the time-scale between points 1 and 2 (Fig 5) can be dramatically longer. This possibility is consistent, at least in part, with a specific version of the panspermia hypothesis. The Hoyle–Wickramasinghe (H-W) theory (Hoyle and Wickramasinghe 1981; Wickramasinghe et al. 2010; Wickramasinghe 2017) suggests that biogenic proto-observing forms, such as viruses, have the cometary origin. Once viruses emerge in comets, the entire cosmos may be seeded with these biogenic structures through the inter-galactic cometary traffic. Thus, the time-scale between points 1 and 2 (Fig 5) may not depend on the processes such as planet formation. Viruses could have emerged very early in the cosmic history in the cometary bodies, as soon as the process of creation of organic elements was completed, in which case the time-scale between points 1 and 2 (Fig 5) could be in the region of several billion years. In support of the H-W theory, it has been argued recently that the retroviruses in the current form originated from the cosmos and coincided with the Cambrian explosion, thus representing the evolutionary driving force (Steele et al. 2018). Furthermore, it is well documented that the human genome is composite, containing DNA from a range of evolutionary ancestors (McFall-Ngai et al. 2013) including 45% of sequences that have the retroviral origin (Moelling and Broecker 2019). However, the H-W theory also assumes that bacteria were created in the interior of comets in which case the time-scale between points 1 and 2 is short (Fig 5).

**5.3 Q3**

Given that the answer to Q2a implies that all life forms are observers (Table 2), it is appropriate to use the term reference classes of observers. Every biological species becomes a reference class. The AP assumption is that the only reference class of observers capable of true observation is *Homo sapiens*. The concept of true observation has three elements: (i) the capacity to sense the cosmic scale information, (ii) the capacity to reflect on the sensory input and (iii) the capacity to send the signal of own existence deep into the cosmos and capture a potentially returning signal (communication). The proposed answer to Q3 is that, apart from *Homo sapiens*, several additional reference classes of observers possess the capacity of true observation (Table 2). These include bacteria, plants and the biosphere as a whole.

This answer may sound controversial or too speculative. However, it is neither less plausible, nor logically inferior to the current thinking behind SETI (see below). In the past, biologists including Ernst Mayr (cited in Lineweaver 2007), expressed deep doubts about the validity of the biological interpretation of SETI. Here is a sketch of the process of true observation which does not depend on the neural-type intelligence. Instead, it is entirely based on the bacterial cognition and interactions between bacteria as cognitive agents and proto-observing units such as viruses.

The first thing to note is that both bacteria and viruses exist as planetary-scale superorganisms – bacteriosphere and virosphere. Sonea and Mathieu (2001) characterized the bacteriosphere as "a world-wide-web of genetic information" that emerged as "a global bacterial superbiosystem" roughly 1 BY since the emergence of first bacteria on Earth. The bacteriosphere is still the most dominant form of life in the biosphere. A recent study suggests that bacteria are, by far, the most abundant life form in the biosphere, far more abundant than the other two domains of life Archaea and Eukarya (Hug et al. 2016). The fact that the most dominant form in the virosphere are phages, or bacterial viruses (Moelling and Broecker 2019), suggests that the bacteriosphere and the virosphere are structurally coupled.

The first SETI assumption is that intelligent observers must be capable of discovering electromagnetic waves. The bacteriosphere "discovered" electromagnetic waves when photosynthetic cyanobacteria emerged in the bacteriosphere. Cyanobacteria can sense the portion of the electromagnetic spectrum between 400 nm and 700 nm (visible light). Thus, the first criterion of the true observation is satisfied – the capacity to sense the cosmic-scale information in the form of bacterial discovery of electromagnetic waves originating from the main star of our planetary system, the Sun (Fig 6; i = "discovery" of electromagnetic waves). The second criterion, or the capacity to reflect on the sensory input, may be interpreted as the product of cyanobacterial sensing of the portion of electromagnetic spectrum combined with its metabolic habit to carry out photosynthesis, the final product of which is the oxygenation of the atmosphere (Fig 6; ii = reflection on the sensory input). Thus, the reflection of the bacterisphere on the sensory input is in line with the biosemiotics concept of a sign reading, relational biology's modelling concept, natural learning as articulated by evolutionary epistemology and autopoiesis as the key process behind the systems view of life (Fig 3; Table 1).

The third criterion within the SETI programme (active SETI), communication or the capacity to send the signal advertising own existence deep into the cosmos and to capture a potentially returning signal, is based on using radio-waves and radio-telescopes, or Van Neuman probes (Tipler 1981). In the case of bacteria, the man-made technology is entirely replaced by a natural technology - a form of biological tropism termed here cosmic tropism (Fig 6; iii = communication). Biological tropism can be defined as the capacity of organisms to produce predictive models about their environments (Louie 2010; see also section 4.1). The cosmic tropism means that a cosmic scale biogenic structure, such as the bacteriosphere, is capable of producing the predictive model of its cosmic environment through advertising itself to the cosmos-wide flow of biogenic particles which may not necessarily be observers, such as viruses - the bacteriosphere is anticipating the virosphere (Fig 6; iii; see also section 4.1). This possibility is conditional on accepting a version of the panspermia hypothesis: either the H-W theory or transpermia. The H-W theory predicts that biogenic particles, including viruses and bacteria, are formed in the interior of comets and that there is a constant flow of biogenic particles from the cosmos towards Earth (Hoyle and Wickramasinghe 1981; Wickramasinghe et al. 2010; Wickramasinghe 2017). Alternatively, the transpermia hypothesis combines the concept of abiogenesis occurring on Earth-like planets with the subsequent transport of microbes, such as viruses and bacteria, to neighbouring planets via rocky ejecta (Davies 2003). Even though the critical evidence for the extraterrestrial origin of life is lacking, both the H-W theory (Steele et al. 2018; 2019) and the transpermia hypothesis (Davies 2003) have been strongly defended. Similarly, the SETI programme lacks critical evidence for the prediction that human-type observers must exist in the universe (Tipler 1981; Lineweaver 2007). Therefore, both the H-W theory/transpermia hypothesis and the SETI programme should be treated equally until the evidence is available to support or reject one or both. By the same token the plant terrestrial biomass, and the biosphere as a whole, may be regarded as the global biogenic structures that, through the process of cosmic tropism, may satisfy the criteria of true observers (Fig 6; iii).

**5.4 Q4**

The consequence of the relationship (structural coupling) between organisms as cognitive agents and their environments, is the emergence of species-centered worlds or Uexküll's *Umwelten* (Uexküll 2010), that form the basis of modern biosemiotics. Similarly, relational biology and the system's view of life interpret the organism-environment coupling as the capacity of organisms to project/anticipate their environments or "bring forth a world" (see Section 4 and Table 1). In the context of Q4, it may be appropriate to call species-centered worlds "cognitive universes" because organisms are cognitive agents structurally coupled with their environments through species-specific sensoria which represent cognitive tool-kits.

The answer to Q4 is that all organisms (observers) possess species-specific observing worlds or collective cognitive spaces, that may be called cognitive universes (Table 2). The cognitive universe of a tick is as functional as the cognitive universe of a human individual (Uexküll 2010). By following on Uexküll's lead, modern biosemiotics interprets the biosphere as a compendium of cognitive universes, or the semiosphere (Hoffmeyer 1996), that we may call the cognitive multiverse. Crossing the boundaries between individual cognitive universes is only possible through a form of biosemiotics known as the cross-kingdom communication (McFall-Ngai et al. 2013). This is apparent, for example, in the case of the communication between bacterial cells in our guts and our brain cells, known as the gut-brain axis (Bruce-Keller et al 2018). Other examples of cross-kingdom communication are discussed elsewhere (Slijepcevic 2018).

Another feature all organisms (observers) share is that each cognitive universe becomes a species-specific civilisation termed here biocivilisations (Table 2). Even though the term civilisation is usually applicable only to the human world, given that the structural coupling between organisms and environments leads to the emergence of natural artefacts including, for example (i) oxygenation of the atmosphere (cyanobacteria), (ii) food products through agriculture (ants and termites) (Mueller et al. 2005), (iii) animal settlements resembling human cities such as underground "cities" of *Atta* and *Acromimex* ants, or termite mounds in Africa and Australia (Wilson 2012), and (iv) huge biogenic structures such as multi-scale natural patterns known as Mima mounds (North America), *murundus* (Brazil) or *heuweltjies* (South Africa) (Tarnita et al. 2017), it is appropriate to consider these natural artefacts byproducts of biological technologies (Slijepcevic 2019c). As a result, the biosphere can be viewed as the global composite biocivilisation. It consists of individual biocivilisations that are structurally coupled to each other. This coupling is termed the interactome (Slijepcevic 2019a), the consequence of which is the biosphere homeostasis (Lovelock and Margulis 1974). (Definitions of cognitive universes, the cognitive multiverse, and biocivilisations, are given in Table 2.)

## 6. Discussion

While the AP concept has been criticised as unscientific (Smolin 2004) or even arrogant (Barash 2018), it has solid support amongst astrophysicists (e.g. Livio and Rees 2005). The position towards AP taken in this study is that AP can be useful, primarily as a method for applying anthropic reasoning to eliminate anthropic bias. The aim of this study was twofold. First, to adapt AP to biology and search for the presence of observing and true observing faculties outside the confines of the human-centered world. Second, to explore the concept of natural intelligence from a new angle that is not biased towards neural-type intelligence.

### 6.1 Observation and true observation

Evidence from different disciplines of biology (section 4; Fig 3; Table1) indicates that the process of cognition is a biological universal. Brandon Carter suggested that the term Cognition Principle represents better what he meant when he was formulating AP (Carter 1983). Carter (1974) and others (Barrow and Tipler 1986) developed AP to understand the position of *Homo sapiens* relative to the rest of the universe. However, Carter (1983) and subsequent interpreters of AP (e.g. Bostrom 2002) did not appreciate enough evidence from biology, which I interpret in this study as a form of anthropic bias. If the capacity of cognition/observation is a biological universal (Section 5.2), several new elements should be added to the AP concept.

First, there are biogenic planetary structures with the cosmic-scale communicative potential (section 5.3) that are usually ignored by mainstream science and philosophy. Such planetary structures are the bacteriosphere and virosphere (Sonea and Mathieu 2001;

Moelling and Broecker 2019). These planetary-scale superbiosystems are best viewed as an invisible planetary cloud that goes to a certain depth into the planetary ground and to a certain height into the planetary atmosphere. The biogenic cloud has the potential to communicate with the biogenic particles coming from the cosmos as predicted by the H-W theory (Wickramasinghe et al. 2010) or transpermia hypothesis (Davies 2003). Thus, the communication reach of the biogenic cloud is truly a cosmic-scale reach, because there is no barrier for viruses coming from a different galaxy (H-W theory), or a different planet (transpermia hypothesis), to make a 'contact'[8] with the body of the Earth-bound bacteriosphere and virosphere. At the heart of this communication capacity is biological tropism which does not require consciousness of the human type, and yet it is based on the principles of cognition (section 5.3). Importantly, the invisible microbial cloud of bacteria and viruses houses all other non-microbial life forms, or macrobes, such as plants and animals (Margulis 1993; 1999; 2004; Sonea and Mathieu 2001). This makes macrobes, including human beings, microbial or bacterial vectors. For example, the International Cosmic Station is already contaminated with bacteria from our microbiome (Checinska et al. 2015). Furthermore, our consciousness-based cognitive faculties and our technology, lag behind the bacterial counterpart in terms of the SETI potential (Fig 6): the human communication window of temporal opportunity is dramatically smaller than that of the bacteriosphere.

      Second, *Homo sapiens* as a macrobe from kingdom Animalia is a composite observer because every single cell in the human body has observing capacities in the sense outlined in section 4. Yet we take for granted the notion that we are biological singletons when it comes to interpreting the concept of observation, even though the ecological collective of our bodies

---

[8] The word 'contact' is used in the mainstream SETI circles to describe the meeting between the Earthbound civilisation and its extraterrestrial counterpart.

houses trillions of cellular observers (Fig 1). This discrepancy between the true individuality and the collective individuality challenges one of the explanatory concepts used to justify the AP style argumentation. The key conundrum of AP, known as the fine-tuning principle, according to which all physical constants of the universe are fine-tuned to the extent that a small change in parameters would mean that life, as we know it, would not emerge at all, is resolved through the concept of physical multiverse (Garriga and Vilenkin 2001; Koonin 2007; Tegmark 2009). According to the model of eternal inflation of the universe, all macroscopic events are repeatable an infinite number of times. Hence, there is not one universe but many universes or the multiverse. The physical multiverse concept is thought to be a scientific concept, meaning that it can be falsified (Tegmark 2009). On the other hand, some aspects of the concept may be too metaphysical to qualify as scientific (Ellis 2011).

However, evidence from biology points towards the existence of the cognitive multiverse (Section 5.4). The human body is a cognitive mini-multiverse: we are conglomerates of viruses, bacteria, archaea, protists, eukaryotic cells housing former bacteria (mitochondria) (McFall-Ngai et al. 2013), culminating in the emergence of the corporate body dominated by the neural-type intelligence and the consciousness (Slijepcevic 2018). Brain and nervous system, required for the rapid assessment of the changing organism-environment interactions that typify life for animals (Musall et al. 2019), combined with the consciousness, may give rise to a potentially misleading impression that we are true biological singletons, or absolute individuals (see Fig 1) completely independent from the rest of the biosphere. The key question then becomes whether the concept of the physical multiverse is an illusion created by the human consciousness, given that the biosphere, containing us as an integrated component, is a complex web of cognitive relationships that cannot be disentangled and "purified" to reflect cognitive spaces of each species.

Carl Gustav Jung divided the human psyche into three areas: individual conscious (ego), individual unconscious and the collective unconscious (Jung 1980). He speculated that the collective unconscious is a biological continuum that can be traced to animals (Jung 1982). However, there is no reason to stop at animals because this violates the continuity thesis of the evolutionary process. Therefore, to bring Jung's thinking in line with the principles of evolution, the collective unconscious can be traced to first life forms, bacteria and archaea. In line with this possibility, recent research implicates the human microbiome as a natural force shaping development of the brain (Dinan et al. 2015), thus opening the route for empirical testing of the concept of collective unconscious all the way down to bacteria and archaea, as first observers on planet Earth from which all other observers descended (Margulis 2004). Here is an interesting question. Is our consciousness falsely projecting the existence of the physical multiverse as a result of the composite nature of our cognitive/observational faculties that (i) originated at the dawn of life with bacteria and archaea and (ii) reflect the entire cognitive multiverse that developed in the last 3.8 BY? If the answer is positive, the AP concept requires deep changes (see also below).

Third, Barrow and Tipler (1986) argued that once an information processing entity emerges in the universe it will never be destroyed. This possibility named FAP (see section 2), has been dismissed outright by some scientists (Gardner 1986). Others tend to interpret it as the emergence of post-biological evolution which may have a greater survival potential than human civilisation (Bostrom 2002). However, both views are ignorant of the planetary-scale information processing entities such as the bacteriosphere, florosphere (the plant terrestrial biomass) and the biosphere as a whole (Slijepcevic 2019a). The bacterial planetary superorganism has been using the natural computation to regulate biogeochemical cycles of organic elements for billions of years (Sonea and Matheiu 2001), a possibility consistent with the informational view of life implicit in the concept of biological determinism (Davies 2003).

Margulis (1999) argued that bacteria are potentially an indestructible form of life. This possibility is supported by the fact that mass extinctions that occurred in the history of life several times have never been able to destroy bacteria. Bacteria have been existing in continuity for at least 3.8 BY, which is most of the existence of planet Earth and more than a quarter of the duration of the entire cosmos. Furthermore, a simple thought experiment supports the notion of bacterial indestructibility[9]. Let us assume that humanity has a pressing need to destroy all bacteria on Earth. Theoretically, we can achieve this. We can use high doses of ionizing radiation, potent DNA-damaging chemicals and other agents capable of destroying life forms. However, this becomes an impossible task. If we attempt to eradicate bacteria from the biosphere we would need to destroy ourselves because we are bacterial vectors – all plants and animals, are carriers of species-specific bacteria-dominated microbiomes (Margulis 1993; Zilber-Rosenberg and Rosenberg 2008). Even if we imagine that an extra-terrestrial civilisation can eradicate bacteria on Earth, this would mean the destruction of the entire biosphere, as the bacteriosphere represents its essential layer. Thus, Barrow and Tipler's FAP potentially exists in the form of the bacteriosphere, but it is wrongly named and interpreted. By contrast, the concept of biological determinism (Davies 2003) allows FAP through the informational context of life (Table 1).

If the above arguments are accepted, the concept of AP requires major changes. In line with Carter (1983) it could be renamed into Cognitive Principle. Similarly, there is a need to update WAP and SAP. SAP was suggested to be a highly teleological concept containing theological overtones. However, when we exclude *Homo sapiens* from the central position it occupies within the AP concept, we end up with the life principle (Davies 2003), cognitive multiverse and biocivilisations (Section 5.4). Even FAP and PAP are not unreasonable in the new interpretation (see above). However, my intention is not to argue that these changes should be accepted. It is enough to point towards the existence of viable explanatory alternatives emerging from the biological research (Section 4) and concepts such as biological determinism (Davies 2003), biocentrism (Lanza and Berman 2009) and informational view of life (Slijepcevic 2019a).

---

[9] It is enough that a small number of bacteria survive to regenerate the bacteriosphere.

**6.2 A new view of intelligence**

Intelligence as a biological trait is almost exclusively interpreted as the human-only capacity to understand the world through the consciousness-based cognition, which can be further enhanced through merging human bodies with AI technologies (Kurzweil 1990; Bostrom 2014; Price 2016). This position is a typical form of anthropic bias. The bias towards the brain, neural intelligence and its derivative, AI, dubbed by a prominent botanist 'brain chauvinism' (Trewavas 2017), is apparent in the AP concept, which ascribes traits of observation and true observation exclusively to *Homo sapiens* (section 2). However, the message of this study, based on the evidence form different disciplines of biology (section 4), is that the process of cognition is a biological universal, rather than the human-only, or higher-animal-only trait. Furthermore, the consciousness of the human type is not a condition for cognition (Lanza and Berman 2009; Trewavas 2017; Calvo et al. 2019).

The simplest way to recognise the status of cognition as a biological universal is to view it as the nature-wide epistemic process, or natural epistemology (section 4). Cognition as a biological universal represents a form of natural epistemology required for organisms, as autonomous natural agents, to adjust to their environments and establish organism-environment interactions which, as epistemological interactions, are subject to the filter of natural selection (Table 1). The standard neo-Darwinian narrative which interprets organisms as passive biological forms, or dumb "lumbering robots", shaped by the interaction between genes and the environment (Dawkins 1976), is challenged by a new interpretation of the organism-environment interactions, known as the extended evolutionary synthesis or "niche construction" theory (Laland et al. 2014), which is in line with the notion of natural epistemology.

The central problem of natural epistemology, provided that we accept it as the theoretical basis behind the nature-wide cognitive process, is how to integrate intelligence in the anthropocentric guise, into a wider evolutionary picture. This problem becomes acute in the case of two research programmes shaped by the anthropocentric interpretation of intelligence, AI (Alexander 2019) and SETI (e.g. the debate between Carl Sagan and Ernst Mayr about the validity of SETI; cited in Lineweaver 2007). Some AI-based predictions clash with the predictions stemming from natural epistemology. A typical example is a prediction of an influential futurist, Ray Kurzweil, according to which the planet Earth will become a gigantic AI-based computer by the year 2099 (Kurzweil 2010). This, according to Kurzweil, is the logical consequence of the emergence of technological singularity, or AI-based superintelligence, which is predicted to occur roughly by the mid 21$^{st}$ century. However, Kurzweil's prediction completely ignores the fact that natural computation on the planetary scale has existed for billions of years in the form of bacterial regulation of biogeochemical cycles of organic elements (Margulis 1999; Sonea and Matheiu 2011).

Similarly, SETI assumes that the only form of intelligence that could exist outside the planet Earth is either the humanity-type intelligence or a higher form of intelligence resulting from the post-biological evolution, e.g. some form of machinocene. However, both options ignore the possibility that the planetary-scale structures, such as the bacteriosphere, can communicate with the biogenic structures potentially existing somewhere else in the cosmos (section 5.3) through the process of biological tropism. It is important to stress that biological tropism is not a passive trait. In line with the four disciplines of biology, the bacteriosphere may be capable of actively anticipating the virosphere, through modelling relations, sign interpretation, natural learning, and autopoiesis (see Section 4).

The problem of integrating anthropic intelligence with natural epistemology is further exacerbated by the additional two factors. First, even in the fields of human cognition and AI the concept of intelligence lacks a unified theoretical basis. For example, at least 70 different definitions of intelligence exist in the literature (Legg et al. 2007). Second, some proponents of plant intelligence interpret it almost like a metaphysical concept that may not be testable by experimental research (Chamovitz 2018).

A necessary step in eliminating anthropic bias when it comes to interpreting the concept of natural intelligence is to recognise that intelligence is not associated exclusively with the human-type cognitive process. The evidence from the four disciplines of biology in this regard (section 4) is overwhelming: the human-type intelligence is only a fragment in the spectrum of natural intelligence (Table 1). However, given existing controversies associated with interpreting human-type intelligence and how AI, as a derivative of this type of intelligence, is integrated with it, a cautious approach is required to fully grasp intelligence as a biological trait. Even though several biologists proposed definitions of intelligence in the evolutionary sense, ranging from integrated problem solving (Treawavas and Baluška 2011), evolutionary fitness (Trewavas 2017; Calvo et al. 2019), information processing as a form of adaptation (Slijepcevic 2018; 2019), etc., these attempts, irrespective of their validity in narrow scientific or philosophical disciplines, cannot be used yet as the basis for the general concept of natural intelligence because of disparity of investigative areas with vested interests and the lack of consensus in the theoretical sense. Instead, a pragmatic approach advocated here is to initiate a wide-ranging discussion between various interested parties including proponents of microbial and plant intelligence, the AI community and its critics, SETI researchers and proponents of panspermia, astrophysicists and astrobiologists, cognitive scientists, information theorists, evolutionary biologists and proponents of the four disciplines of biology, to establish a coherent theoretical and experimental platform for the study of intelligence in the biological and evolutionary sense. The collective debate may be beneficial to all parties involved. A memorable debate between Carl Sagan and Ernst Mayr (cited in Lineweaver 2007) that took place more than twenty years ago is a good example of a constructive presentation of different views, which resulted in a more mature interpretation of SETI.

**Table 1.** Concepts of organisms, organism-environment interactions, information and information processing from the perspective of four disciplines of biology.

|  | **Organisms as autonomous natural agents** | **Structural coupling organism-environment** | **Information (environmental stimulus)** | **Information processing** |
|---|---|---|---|---|
| **Relational biology** | Anticipatory systems (Rosen 1985) | Modelling relationship - organisms possess internal models of themselves and of their environments (Rosen !985, 1991; see also Fig 2) | Structural and functional information (Kineman 2007) | Encoding and decoding (Fig 2). Structure = model encoding; function = model decoding (Kineman 2007) |
| **Evolutionary epistemology** | Cognitive agents (Slijepcevic 2019a) | Natural learning (Plotkin 1982; Slijepcevic 2019a) | Latent information ($I_p$) and control information ($I_c$) (Corning 2007; Slijepcevic 2019a) | Natural learning algorithm (Plotkin 1982; Slijepcevic 2019a) |
| **Biosemiotics** | Sign interpreters or semiosis users (Kilstrup 2015) | Sign utilization (Kilstrup 2015): a triadic relationship expressed as Ψ (O, R, I) where O=object, R=representamen, and I=interpretant. | Semiotic scaffolding (Hoffmeyer 2015) | Sign establishment phase and a sign interpretation phase (Kilstrup 2015) |
| **Systems view of life** | Autopoietic units (Maturana and Varela 1980) | The process of embodied cognition (Capra and Luisi 2014) | "Difference that makes a difference" (Bateson 1979) | Autopoiesis and self-organization (Maturana and Varela 1980) |

**Table 2.** Answers to the four questions asked in Section 3.

| Question | Answer |
|---|---|
| **Q1**: Are human observers the only observers within the pool of terrestrial life forms? | Human observers are not the only observers within the pool of terrestrial life forms. |
| **Q2a**: If not, what's the frequency of non-human observers within the pool of terrestrial life forms (some life forms, or all life forms)? | All forms of life, from bacteria to animals, are observers. |
| **Q2b**: If all life forms are observers, what's the boundary between the observing and non-observing capacities? | The boundary between observing and non-observing life forms are biogenic proto-observing forms such as viruses. |
| **Q3**: Are there true observers within the pool of terrestrial life forms amongst the reference classes of observers that are not human? | Yes, there are true observers within the pool of terrestrial life forms that are not human observers. These include the bacterial planetary superorganism (bacteriosphere), the plant terrestrial biomass (florosphere) and the biosphere as a whole. |
| **Q4**: Do human observers and other observers and true observers share common features? | The common features all observers and true observers share include cognitive universes and biocivilisations. The cognitive universe is the collective cognitive space of a species. The biocivilisation is the collective phenotype emerging from the cognitive universe. The biosphere thus becomes the cognitive multiverse, or the composite global biocivilisation. |

**Legend for figures**

**Figure 1.** *Mona Lisa* (c.1503-1506) and *Flora* (1588). Images courtesy of Wikipedia.

**Figure 2.** A. The modelling relationship between the natural system (NS) and the formal system (FS) according to Rosen (1991). B. Inclusion of structure and function in the modelling relationship (Kineman et al. 2007). For details see the text. Illustrations are adapted from Rosen (1991) and Kineman et al. (2007).

**Figure 3.** The three-component process of cognition from the perspective of four disciplines of biology.

**Figure 4.** The distinction between observers (all organisms from bacteria to animals) and proto-observers (viruses) relative to the genome size. The transition zone between the living (observers) and non-living (proto-observers) is represented by giant viruses such as mimiviruses. Mimiviruses possess almost all components required for independent living. The transition zone is continuous; the use of the separating line is for illustration purposes. *P. ubique* is the smallest living bacterium. Adapted from Moelling and Broecker (2019) (Figure 3 in their open access article distributed under the terms of a CC-BY license; modification consists of adding two words and a line in red on top of figure).

**Figure 5.** The timeline of the emergence of organisms or true observers. The universe in its early stages lacks observers. A pre-requisite for the emergence of proto-observers and observers is the set of organic elements required for their function. As soon as the set of required elements is formed, this enables the emergence of proto-observers (point 1). Observers emerge (point 2) either very soon after the emergence of proto-observers (as in the case of (i) abiogenesis or (ii) tranpsermia; see text for details) or long after the emergence of proto-observers (as in the case of H-W theory; see text for details). Drawing not to scale.

**Figure 6.** The concept of true observers applied to the bacteriosphere. For the description of three elements, i-iii, required for the concept of true observation, see the text.

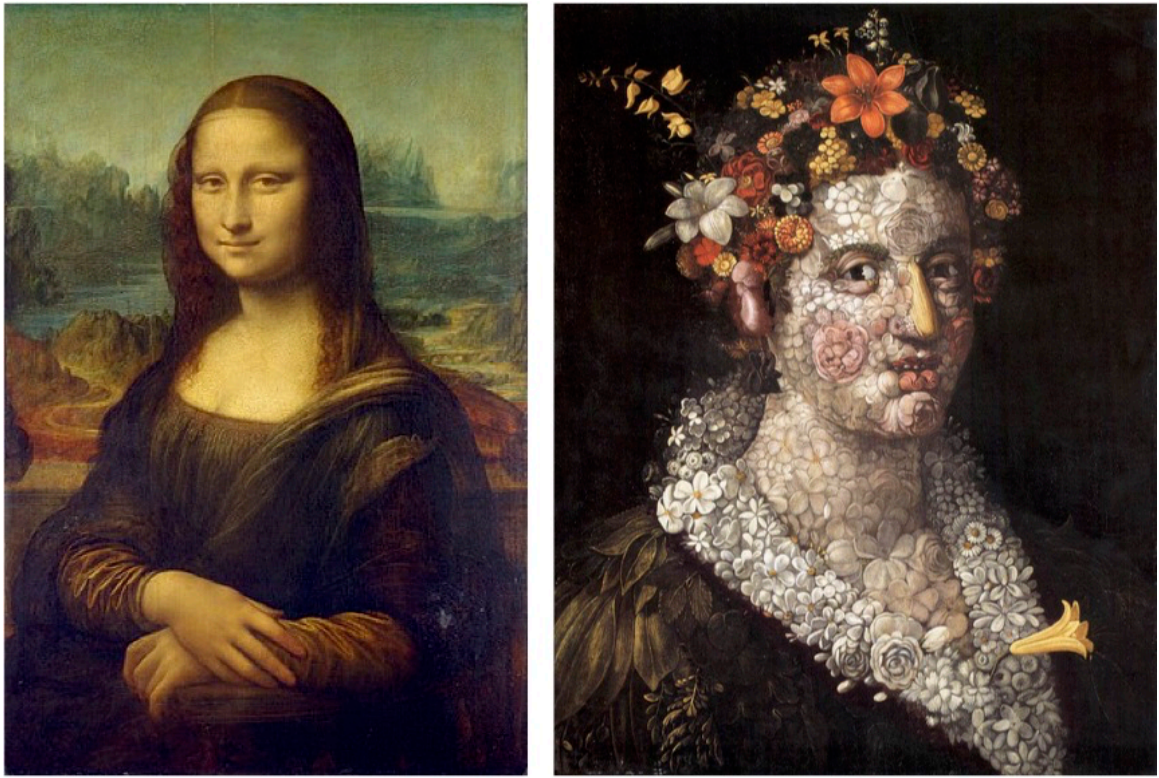

Figure 1

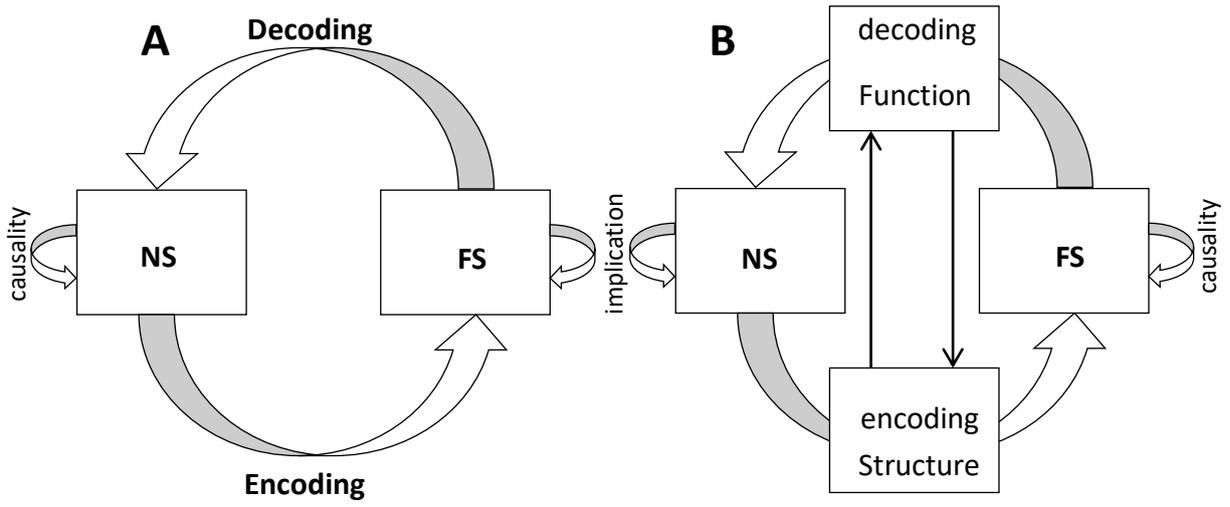

Figure 2

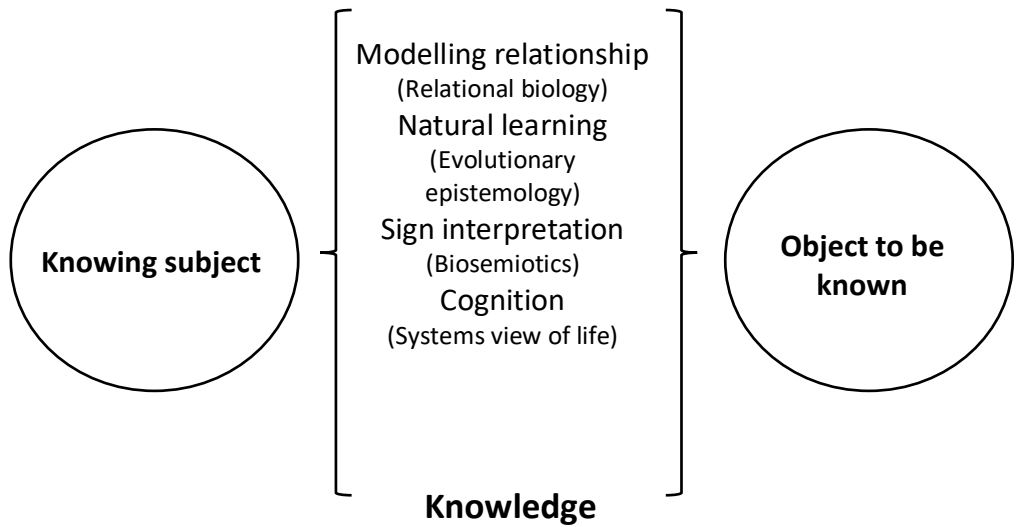

Figure 3

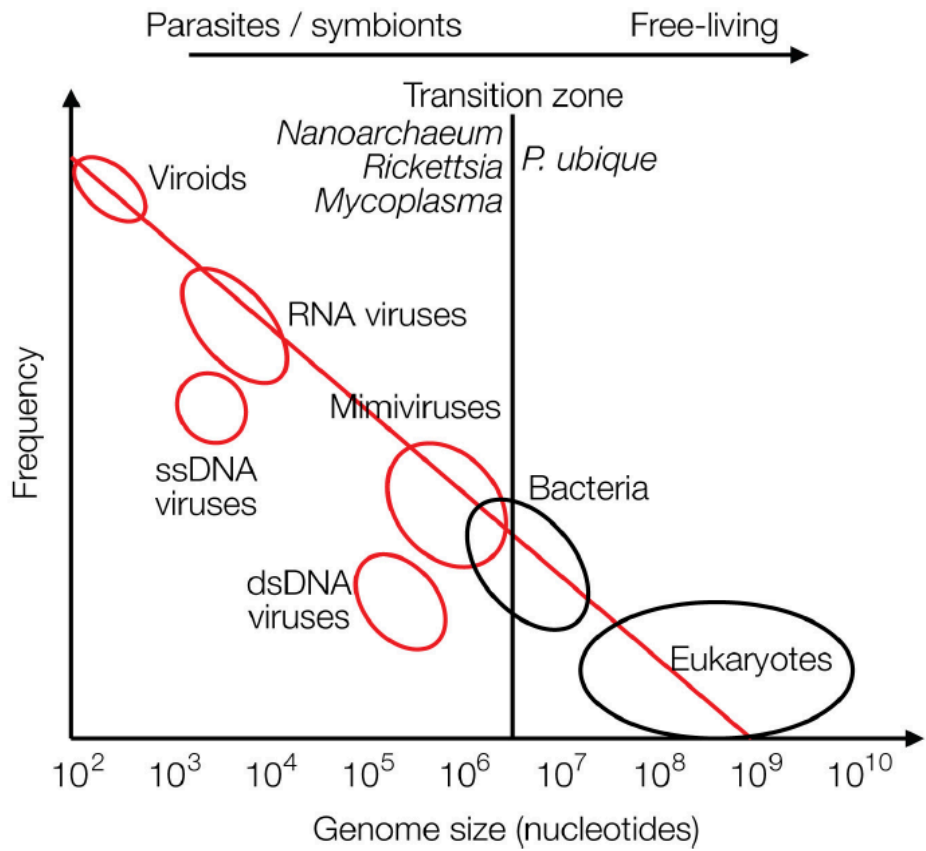

Figure 4

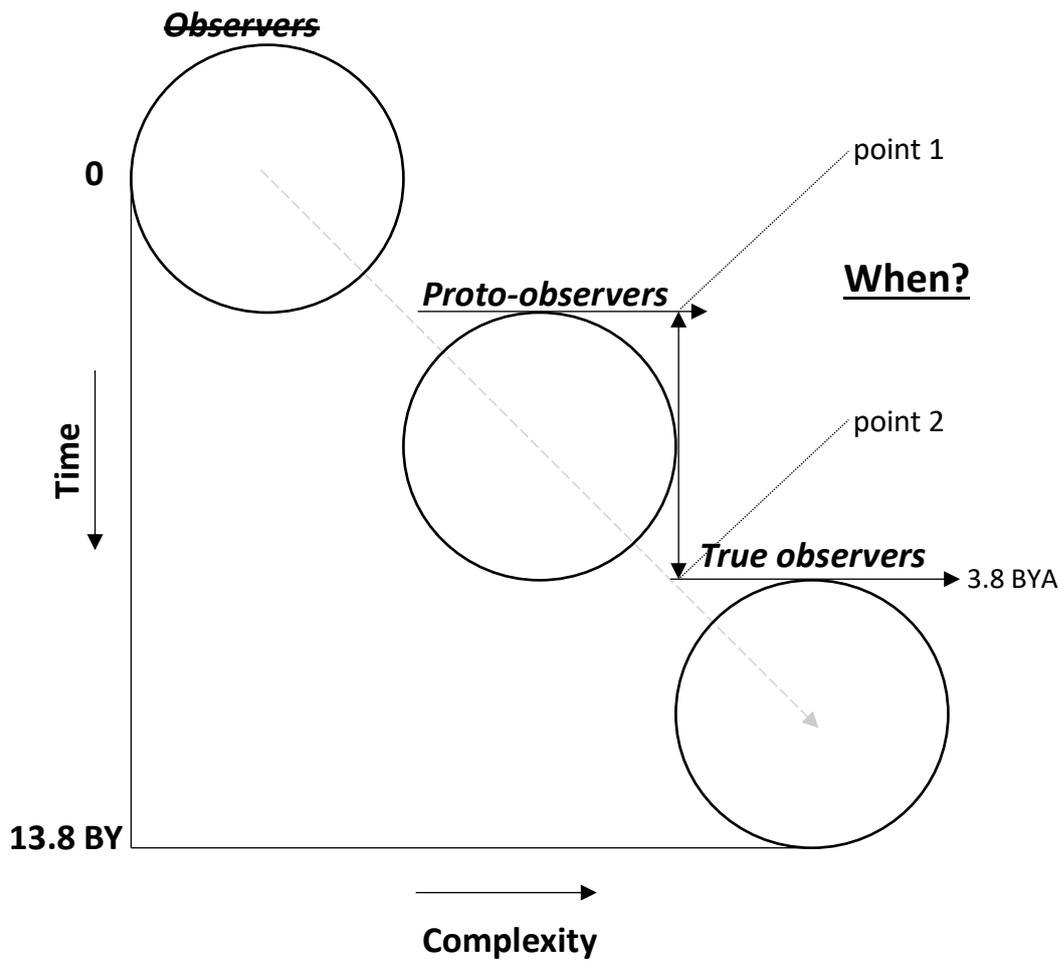

Figure 5

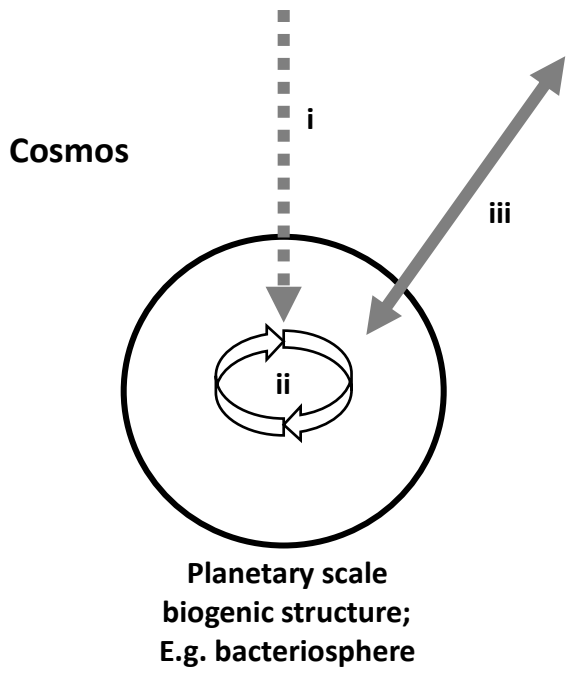

Figure 6